\documentclass[utf8]{frontiersFPHY} 

\usepackage{url,hyperref,lineno,microtype,subcaption}
\usepackage[onehalfspacing]{setspace}
\usepackage{epsfig} 
\usepackage{graphics}
\usepackage{natbib}

\def\keyFont{\fontsize{8}{11} \helveticabold }
\def\firstAuthorLast{Del Moro {et~al.}} 
\def\Authors{Agnese Del Moro\,$^{1,*}$, David~M.~Alexander\,$^{2}$, Franz~E.~Bauer\,$^{3,4,5,6}$, Emanuele Daddi\,$^{7}$, Dale D. Kocevski\,$^{8}$, Flora Stanley\,$^{9}$ and Daniel H. McIntosh\,$^{10}$}

\def\Address{$^1$ Max-Planck-Institut f{\"u}r Extraterrestrische Physik (MPE), Garching, Germany \\
$^2$ Centre for Extragalactic Astronomy, Department of Physics, Durham University, Durham, UK\\
$^3$ Instituto de Astrof\'{\i}sica, Facultad de F\'{i}sica, Pontificia Universidad Cat\'{o}lica de Chile, Santiago, Chile\\
$^4$ Millenium Institute of Astrophysics, Santiago, Chile\\
$^5$ EMBIGGEN Anillo, Concepci\'{o}n, Chile\\
$^6$ Space Science Institute, Boulder, Colorado, USA\\
$^7$ CEA-Saclay, France\\
$^8$ Department of Physics and Astronomy, University of Kentucky, Lexington, Kentucky, USA \\
$^9$ Department of Space Earth and Environment, Chalmers University of Technology, Onsala Space Observatory, Onsala, Sweden \\
$^{10}$ Department of Physics \& Astronomy, University of Missouri-Kansas City, Kansas City, Missouri, USA}

\def\corrAuthor{Dr Agnese Del Moro}

\def\corrEmail{adelmoro@mpe.mpg.de}

\renewcommand\bibname{{\footnotesize References \\[-1cm]}}
\def\aj{AJ}%
\def\araa{ARA\&A}%
\def\apj{ApJ}%
\def\apjl{ApJ}%
\def\apjs{ApJS}%
\def\aap{A\&A}%
\def\aapr{A\&A~Rev.}%
\def\mnras{MNRAS}%
\def\nar{New A Rev.}%
\def\nat{Nature}%
\begin{document}
\onecolumn
\firstpage{1}

\title[Luminous, heavily-obscured infrared QSOs]{Luminous and obscured quasars and their host galaxies} 

\author[\firstAuthorLast ]{\Authors} 
\address{} 
\correspondance{} 

\extraAuth{}

\maketitle

\begin{abstract}
The most heavily-obscured, luminous quasars might represent a specific phase of the evolution of actively accreting supermassive black holes and their host galaxies, possibly related to mergers. We investigated a sample of the most luminous quasars at $z\approx1-3$ in the GOODS fields, selected in the mid-infrared band through detailed spectral energy distribution (SED) decomposition. The vast majority of these quasars ($\sim$80\%) are obscured in the X-ray band and $\sim$30\% of them to such an extent, that they are undetected in some of the deepest (2 and 4 Ms) {\it Chandra} X-ray data. 
Although no clear relation is found between the star-formation rate of the host galaxies and the X-ray obscuration, we find a higher incidence of heavily-obscured quasars in disturbed/merging galaxies compared to the unobscured ones, thus possibly representing an earlier stage of evolution, after which the system is relaxing and becoming unobscured.

\tiny
 \keyFont{ \section{Keywords:} galaxies: active, quasars: general, quasars: supermassive black holes, X-rays: galaxies, infrared: galaxies, galaxies: star formation} 
\end{abstract}

\section{Introduction}

The similarity between the accretion history of galaxies and supermassive black holes (SMBHs), peaking at redshift $z\approx1-2$ (e.g., \cite{madau1996, hopkins2006a, brandt2015}), suggests that there is a connection between the evolution of a galaxy and the black hole in their center. Such connection has also been hinted by the observed correlations between the BH mass and the velocity dispersion of the stars in the bulge ($M_{\rm BH}-\sigma$ relation; \cite{ferrarese2000,gebhardt2000}) or with the bulge mass ($M_{\rm BH}-M_{bulge}$; \cite{kormendy1995, magorrian1998}). Whether this parallel evolution is simply due to a larger gas supply at high redshift, feeding both the SMBH and star formation (SF), or whether there are other processes self-regulating the SMBH and galaxy growth {(e.g., AGN feedback)} is still uncertain (e.g., \cite{alexander2012, kormendy2013}).

Studying active galactic nuclei (AGN) at all cosmic epochs is crucial to fully understand the accretion history of the SMBHs and their role in galaxy evolution. However, most of the accretion onto SMBHs is expected to be heavily obscured by dust and gas, making the identification of the most obscured AGN population very challenging, even in the deepest X-ray surveys. {According to the unified model (e.g. \cite{antonucci1993}), AGN appear obscured due to orientation effects, when our line-of-sight crosses high column densities ($N_{\rm H}$) of circumnuclear material, the so called ``torus''. However, it has been suggested that the most heavily-obscured, Compton-thick (CT; where $N_{\rm H}>1.5\times10^{24}$~cm$^{-2}$) AGN}, especially the (intrinsically) most luminous ones, could {represent} a particular phase of galaxy evolution, associated to a major merger, when a lot of gas and dust are funnelled {into} the centre of the galaxy, {deeply} hiding the active nucleus within it (\cite{dimatteo2005, hopkins2006a}; see also \cite{alexander2012}, for a review). Yet, the emission reprocessed by the obscuring dust is re-emitted in the mid-infrared (MIR) band, which can therefore be used to find even the most obscured and elusive quasars.

\section{Sample selection}\label{sample}

The sample was selected from a large catalog of 24~$\mu$m-detected sources within the GOODS-{\it Herschel} North and South fields. We performed detailed SED decomposition using {\it Spitzer} 8, 16 and 24~$\mu$m and {\it Herschel} 100, 160 and 250~$\mu$m data, to separate the AGN from SF emission. We adopted the AGN and star-forming galaxy (SFG) templates described in \cite{mullaney2011} and \cite{delmoro2013}; details of the SED fitting are described in \cite{delmoro2016}. Amongst these sources we selected the most luminous quasars in the mid-infrared (MIR) band, with rest-frame 6~$\mu$m luminosity of $\rm \nu L_{\rm AGN,6\mu m}>6\times10^{44}$~erg~s$^{-1}$, corrected for the galaxy contribution, at redshift $z=1-3$. This selection results in a sample of 33 sources.  

\section{Analyses and results}

\subsection{AGN: heavily obscured population at $z\approx2$}

To characterise the quasars in our sample, we used the deep X-ray {\it Chandra}\ data available in the {\it Chandra} Deep Field North (CFD-N; 2~Ms; \cite{alexander2003}) and {\it Chandra} Deep Field South (CDF-S 4~Ms; \cite{xue2011}). {For details on the data reduction we refer to \cite{alexander2003, luo2008, xue2011}.} Of our 33 quasars, 24 ($\sim$73\%) are detected in the X-ray band, while 9 ($\sim$27\%) remain undetected, despite being intrinsically very luminous in the MIR band. These sources are candidates to be the most heavily obscured, CT AGN. 

\begin{figure}[!t]
\begin{center}
\includegraphics[width=9cm]{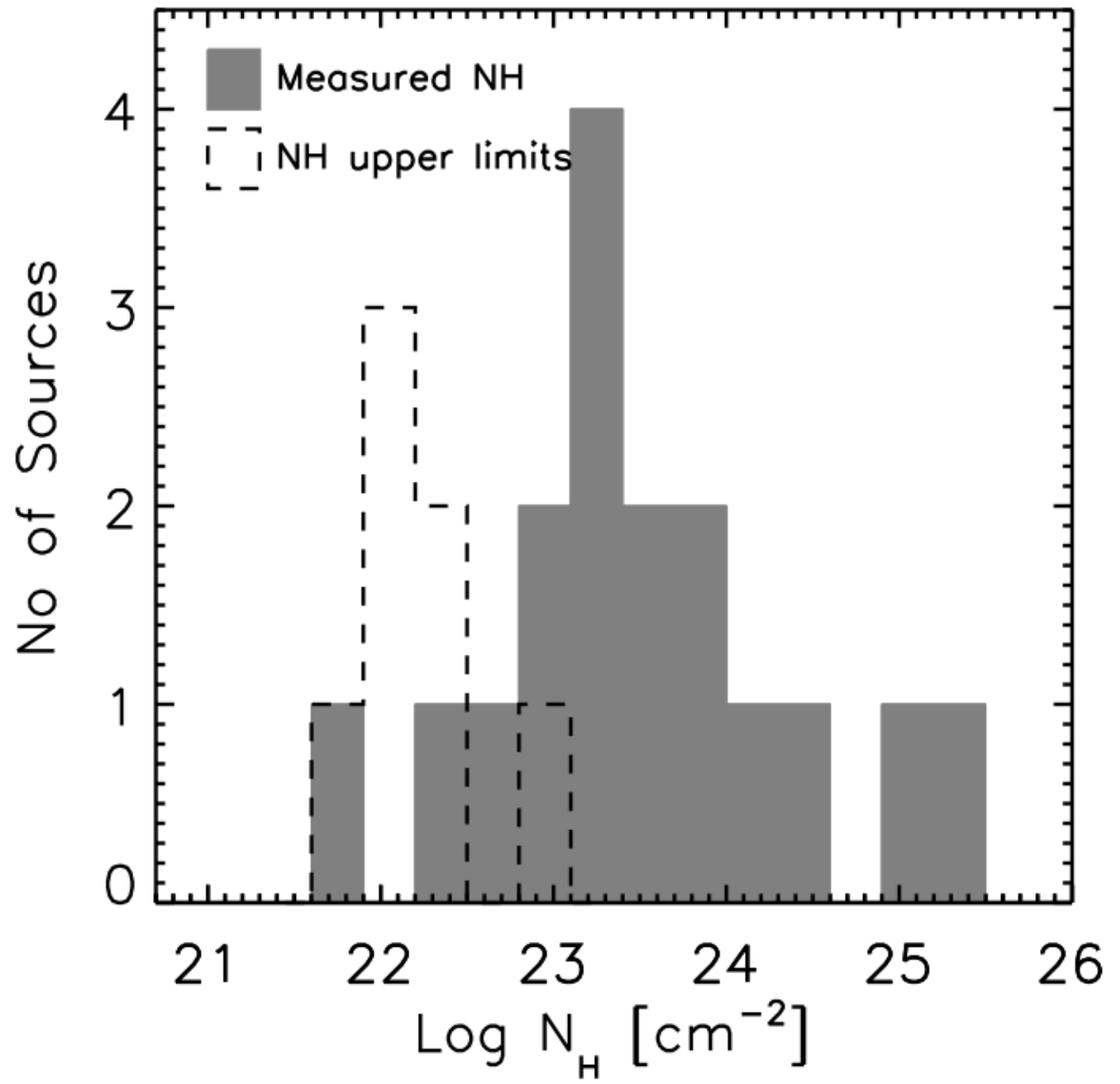}
\end{center}
\caption{Distribution of the X-ray column density ($N_{\rm H}$) for the X-ray-detected MIR quasars ($\sim$70\% of the sample). The dashed histogram represents the $N_{\rm H}$ upper limits (calculated at a 90\% confidence level).}\label{fig:1}
\end{figure}

For the sources that are detected in the X-rays, we extracted {the spectra using {\it ACIS Extract} (AE; \cite{broos2010, broos2012})} and analysed {them} using a simple absorbed power-law model {(including Galactic and intrinsic absorption)} to constrain the amount of $N_{\rm H}$. In Fig. \ref{fig:1} we show the $N_{\rm H}$ distribution of the sample. We find that the majority of these quasars (16/24; $\sim$67\%) are obscured by columns of $N_{\rm H}>10^{22}$~cm$^{-2}$, of which more than half (9/16) are heavily obscured ($N_{\rm H}>2\times10^{23}$~cm$^{-2}$). Amongst these heavily-obscured sources, we can identify six of them as CT quasars from the X-ray spectral analysis, using spectral models appropriate for heavily-obscured sources, such as, PLCABS (\cite{yaqoob1997}) and TORUS (\cite{brightman2011}). The fraction of obscured quasars in our sample reaches $\sim$76\%, and 54\% of heavily-obscured quasars, if we include the X-ray-undetected sources, assuming that these are the most heavily CT ones. Indeed, the comparison between the intrinsic luminosity at 6~$\mu$m and the X-ray luminosity upper limit of these sources suggests that the X-ray emission is heavily suppressed compared to the intrinsic $L_{\rm X}-L_{6~\mu m}$ relation found for AGN (e.g. \cite{lutz2004, fiore2009, gandi2009}), making them very good candidates to be heavily CT quasars. 

We note that amongst the X-ray undetected quasars in the sample, there is one source, \#28 (see Table 1 from \cite{delmoro2016}), that is now detected in the 7~Ms CDF-S catalog (XID 28; \cite{luo2017}). This source has a very flat effective photon index of $ {\Gamma <0.93}$ {(compared to the typical $\Gamma\approx1.8$ for unabsorbed AGN)} and an extremely low rest-frame X-ray luminosity ($L_{\rm 0.5-7~keV}\approx1.5\times10^{41}$~erg~s$^{-1}$, uncorrected for absorption) {compared to its intrinsic luminosity measured in the MIR band (${log~\nu~L_{\rm 6~\mu m}=45.97}$~{erg~s$^{-1}$}),} consistent with the upper limit reported in our analysis (\cite{delmoro2016})\footnote{We note however, that the photometric redshift assumed in our analysis of this source ($z=2.55$) differs from that reported in the Luo et al. 2017 (\cite{luo2017}) catalog ($z=1.81$).}, supporting our assumption that the source might be a heavily obscured, CT quasar.  

These results suggest that there is a large population of heavily-obscured, intrinsically luminous quasars at high redshift, which are very elusive even for deep X-ray surveys. These sources might constitute a special phase of the BH-galaxy evolution, where the actively growing BH is embedded in large amounts of gas and dust, possibly as a result of a recent merger.  

\subsection{AGN host galaxies: star-formation rates and merger fraction}

\begin{figure}[t]
\begin{center}
\includegraphics[width=9cm]{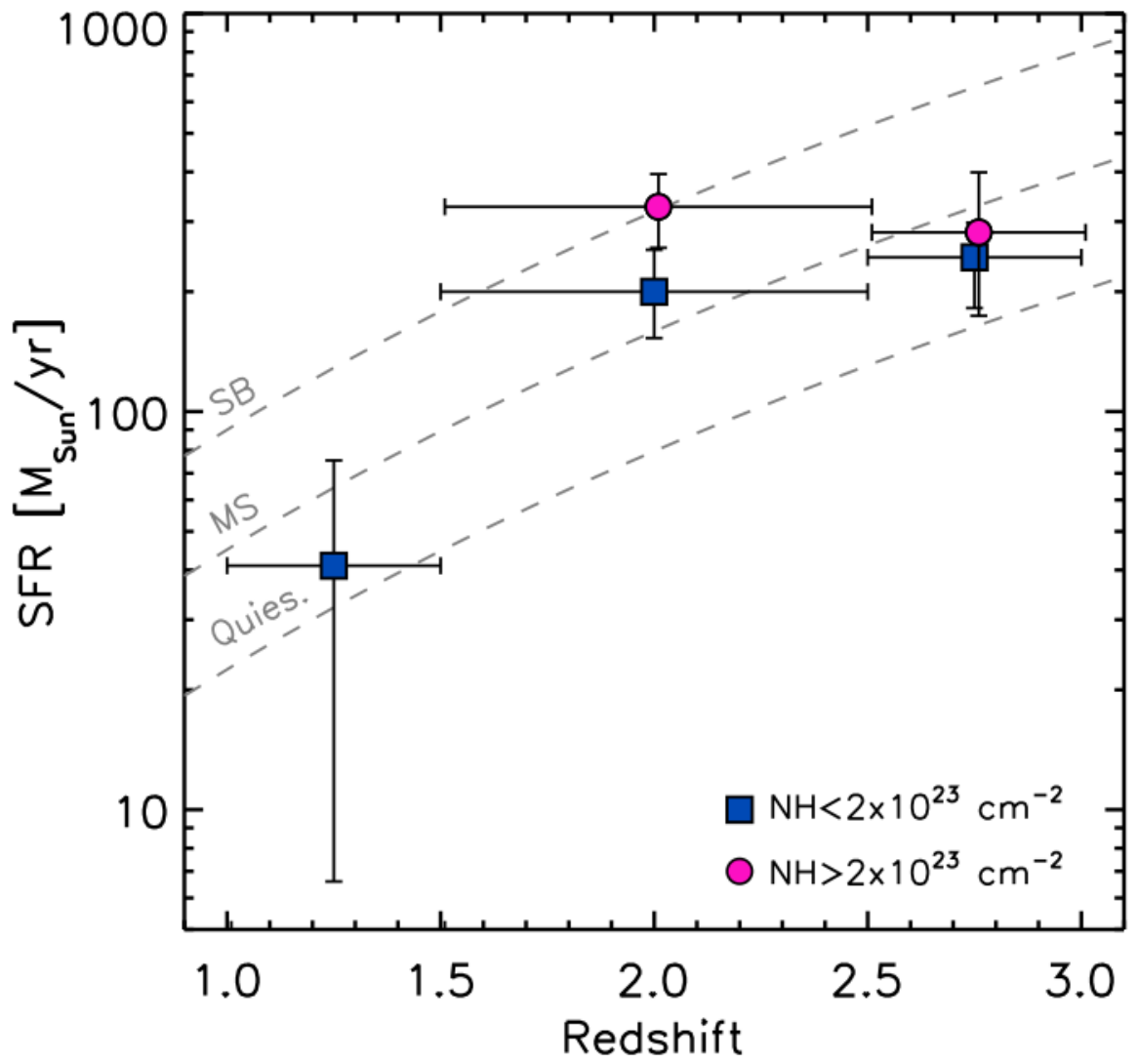}
\end{center}
\caption{Average star-formation rate (SFR) for the unobscured/moderately-obscured quasars ($N_{\rm H}<2\times10^{23}$~cm$^{-2}$; blue squares) and the heavily-obscured quasars ($N_{\rm H}>2\times10^{23}$~cm$^{-2}$; magenta circles) in our sample, divided in three redshift bins. {None of heavily-obscured quasars are detected in the lowest redshift bin.} The grey dashed lines represent the SFR track for main-sequence (MS), starburst (SB; SFR$_{MS}\times2$) and quiescent (Quies.; SFR$_{MS}/2$) galaxies with a typical mass of $M_*=9\times10^{10}$ {$ M_{\odot}$} (e.g., \cite{elbaz2011}).}\label{fig:2}
\end{figure}

To study the characteristics of the host galaxies of these MIR-luminous quasars, we investigated their star-formation rates (SFRs) and their morphologies, in particular the disturbance or distortion of their morphology, as an indication of galaxy mergers and interactions. We derived the SFR of each galaxy from its far-infrared (FIR) luminosity (or upper limit) resulting from the SED decomposition in the IR band (see Sect. \ref{sample} and \cite{delmoro2016}), assuming a Salpeter initial mass function (\cite{salpeter1955}) and the relation from Kennicutt (1998; \cite{kennicutt1998}). We also calculated the average SFR separately for the unobscured/moderately obscured quasars ($N_{\rm H}<2\times10^{23}$~cm$^{-2}$) and for the heavily-obscured quasars ($N_{\rm H}>2\times10^{23}$~cm$^{-2}$) in three different redshift bins (see Fig. \ref{fig:2}). To estimate the average SFRs accounting for the upper limits, we used the Kaplan-Meier (KM) product limit estimator (\cite{Kaplan1958}), a non-parametric maximum-likelihood estimator of the distribution function (see \cite{stanley2015}, for details on the method). 

We find that the average SFR increases with redshift, in general agreement with the SFR main sequence of galaxies (Fig. \ref{fig:2}). Moreover, although the heavily-obscured sources seem to have a slightly enhanced SFR compared to the unobscured/moderately obscured ones (especially at $z\approx2$), these differences are not statistically significant, and therefore we find no clear dependence of the amount of SF in the galaxy with the X-ray obscuration of the quasar. This suggests that the obscuration in the X-ray band is likely confined in the nuclear regions and not related to the presence of gas on larger scales (e.g., \cite{rosario2012, rovilos2012}).

\begin{figure}[t]
\begin{center}
\includegraphics[width=9cm]{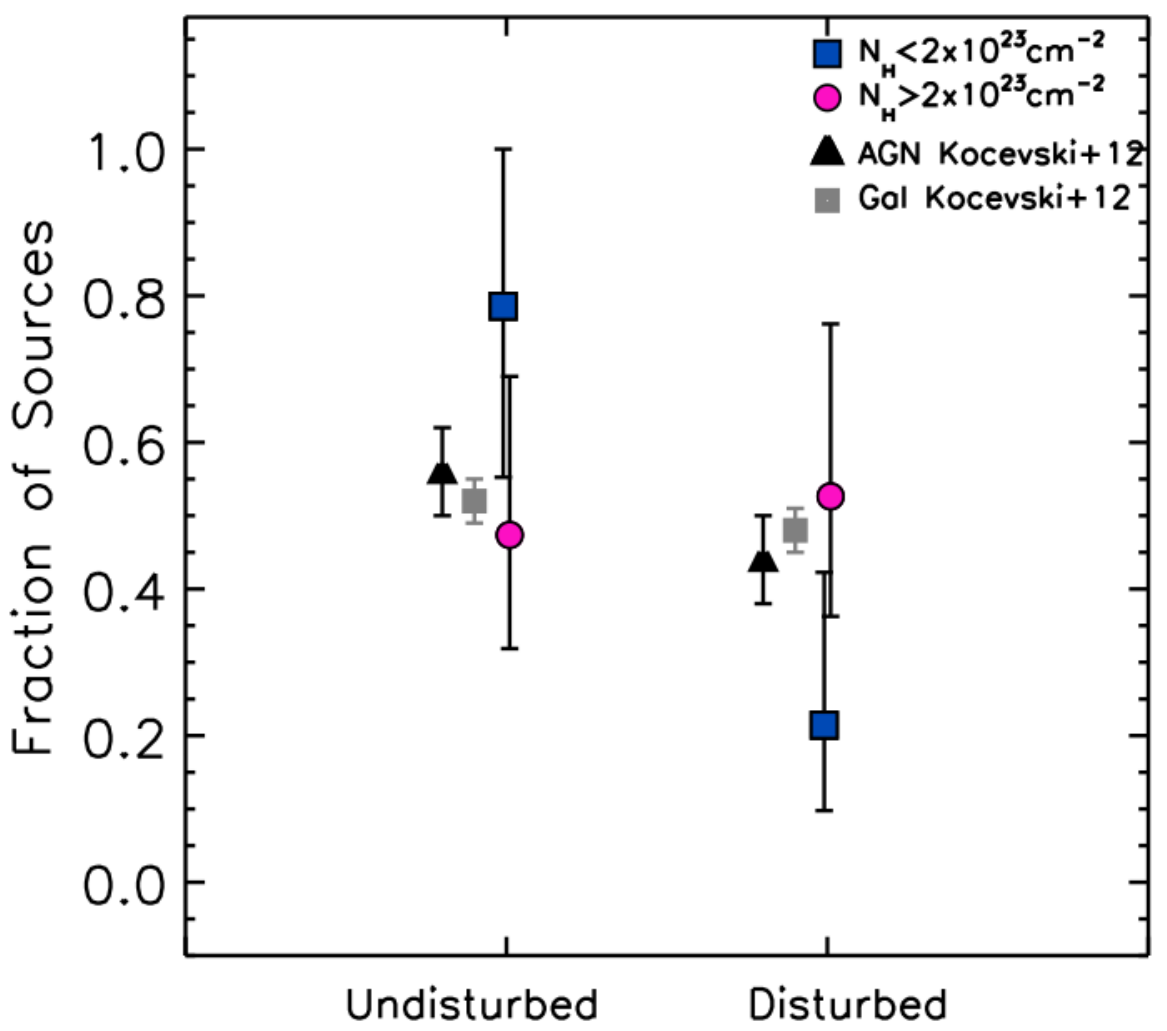}
\end{center}
\caption{Fraction of IR quasar hosts showing disturbed and undisturbed galaxy morphologies, classified using HST images. The fractions and 1$\sigma$ uncertainties for the unobscured/moderately-obscured quasar hosts ($N_{\rm H}<2\times10^{23}$~cm$^{-2}$) are plotted as blue squared and for the heavily-obscured quasars ($N_{\rm H}\ge2\times10^{23}$~cm$^{-2}$) as magenta circles. For comparison we also show the fractions for $z\approx2$ AGN and non-AGN samples from \cite{kocevski2012}: filled black triangles and grey squares, respectively. The unobscured/moderately-obscured quasars reside preferentially in undisturbed systems, while the heavily-obscured quasars are equally found in disturbed and undisturbed systems.}\label{fig:3}
\end{figure}

Using the high-resolution optical HST images available in the GOODS-N and GOODS-S fields, as part of the GOODS and CANDELS projects (\cite{giavalisco2004, grogin2011}), we visually inspected the morphology of these galaxies to identify signs of distortions or disturbances, which would indicate a recent galaxy merger/interaction event. We adopted a similar classification scheme to that used by \cite{kocevski2012} (see \cite{delmoro2016}, for details), separating the sources into ``disturbed'' and ``undisturbed''. In Fig. \ref{fig:3} we show the fraction of sources that have disturbed and undisturbed morphologies over the total, dividing them again into unobscured/moderately obscured (blue squares) and heavily obscured (magenta circles), as in Fig. \ref{fig:2}. We find that a relatively high fraction of our sources shows signs of distortions/interactions ($\approx$40\%), higher than those typically found at low redshift ($\sim$15-20\% at $z<1$; e.g., \cite{cisternas2011}). This is in agreement with the trend of increasing major-merger fraction with redshift seen in previous works (e.g., \cite{conselice2003, treister2006, kartaltepe2007}). We find that, on average, the most-heavily obscured quasars tend to have more disturbed morphologies than the unobscured/moderately obscured ones ($\approx$53\% vs. $\approx$20\%, respectively); although the errors on these fractions are large due to the small number of sources in our sample, the difference between the two quasar populations is significant at the 90\% confidence level (Fisher exact probability test: $P=0.087$). This trend is seen also in other studies (e.g., \cite{kocevski2015, ricci2017}). We note, however, that these studies investigated samples in different redshift and/or luminosity ranges compared to ours, and therefore the actual fractions of sources with disturbed morphologies are not directly comparable. The smaller fraction of disturbed systems we found for the unobscured/moderately-obscured quasars could be interpreted within the SMBH-galaxy evolutionary models, as the unobscured quasars would represent a later stage of the evolution compared to the heavily-obscured sources and the distortion features due to mergers or interactions might have faded by the time these quasars are observed as unobscured, given the relaxation time of a galaxy is typically $\sim200-400$~Myr (e.g., \cite{lotz2010}). On the other hand, for the most heavily-obscured quasars, which might represent a younger stage of evolution after a merger in this scenario, the signatures of the recent interactions are still evident in their hosts.

\section{Conclusions}

We have investigated the AGN and host galaxy properties of a sample of 33 quasars at $z=1-3$ within the GOODS-{\it Herschel} fields, selected in the MIR band through detailed SED analysis to have an intrinsic AGN luminosity of ${\rm \nu L_{\rm 6~\mu m}}>6\times10^{44}$~erg~s$^{-1}$. Despite being intrinsically the most luminous quasars within these fields, $\sim$26\% of them are not detected in the deep 2 and 4~Ms {\it Chandra} X-ray data covering this sky area. 

We performed X-ray spectral analysis of the 24 X-ray-detected sources to investigate the AGN properties, and we found that the vast majority ($\sim$67\%; 16/24 sources) are obscured by $N_{\rm H}>10^{22}$~cm$^{-2}$, with more than half of them (9/16) being heavily obscured ($N_{\rm H}>2\times10^{23}$~cm$^{-2}$). Including the X-ray undetected sources, which are likely to be the most heavily CT AGN, these fractions reach $\sim$76\% ($\sim$54\% are heavily obscured). This means that there is a very large population of heavily obscured, intrinsically luminous quasars at redshift $z\approx2$, which can be revealed in the IR band, but remains (in part) undetected in the X-ray band. 

We investigated the host galaxy properties of these quasars through their SFR, measured in the FIR band from SED fitting using {\it Spitzer} and {\it Herschel} data, and did not find any strong link between the amount of SF and the X-ray obscuration, possibly suggesting that the X-ray obscuration is mostly concentrated in the nuclear regions and does not depend on the presence of gas on larger scales.

We also visually classified the morphology of these quasars as disturbed or undisturbed using high-resolution HST data to identify signs of distortions/asymmetries in the galaxies. We find that a significant fraction ($\sim$40\%) have disturbed morphologies, suggesting they have experienced a recent merger or interaction event. We find a larger fraction of sources with disturbed morphologies amongst the heavily-obscured quasars ($\sim$53\%) rather than the unobscured/moderately-obscured ones ($\sim$20\%). Our results possibly support the SMBH-galaxy evolutionary scenario where the heavily-obscured quasars represent an earlier stage of evolution after the merger, while the unobscured quasars represent a later stage of the evolution, when the system has relaxed, the signs of interaction have already faded, and the nucleus becomes unobscured.

\section*{Conflict of Interest Statement}

The authors declare that the research was conducted in the absence of any commercial or financial relationships that could be construed as a potential conflict of interest.

\section*{Author Contributions}

ADM developed the concept, performed all the analyses and wrote the manuscript. DMA initiated and helped developing the concept. FEB provided the X-ray spectra. ED compiled the multiwavelength catalog, which was used for the SED analysis. DDK and DHM provided the optical HST images, used to classify the galaxy morphology. FS calculated the mean star-formation rates.

\section*{Funding}
This research was supported by the UK Science and Technology Facilities Council (STFC, ST/L00075X/1); CONICYT-Chile (Basal-CATA PFB-06/2007, ``EMBIGGEN'' Anillo ACT1101, FONDECYT Regular 1141218); the Ministry of Economy, Development, and Tourism's Millennium Science Initiative through grant IC120009,  awarded to The Millennium Institute of Astrophysics (MAS).


\end{document}